\documentclass[11pt]{article}
\usepackage[]{amsmath,amssymb}
\usepackage{graphics,epsfig}

\begin{document}
\date{}
\title{Mutual information challenges entropy bounds}
\author{H. Casini\footnote{e-mail: casini@cab.cnea.gov.ar} \\
{\sl Centro At\'omico Bariloche,
8400-S.C. de Bariloche, R\'{\i}o Negro, Argentina}}
\maketitle

\begin{abstract}
We consider some formulations of the entropy bounds at the semiclassical level. The entropy $S(V)$ localized in a region $V$ is divergent in quantum field theory (QFT). Instead of it we focus on the mutual information $I(V,W)=S(V)+S(W)-S(V\cup W)$ between two different non-intersecting sets $V$ and $W$. This is a low energy quantity, independent of the regularization scheme. In addition, the mutual information is bounded above by twice the entropy corresponding to the sets involved. Calculations of $I(V,W)$ in QFT show that the entropy in empty space cannot be renormalized to zero, and must be actually very large. We find that this entropy due to the vacuum fluctuations violates the FMW bound in Minkowski space. The mutual information also gives a precise, cutoff independent meaning to the statement that the number of degrees of freedom increases with the volume in QFT. If the holographic bound holds, this points to the essential non locality of the physical cutoff. Violations of the Bousso bound would require conformal theories and large distances. We speculate that the presence of a small cosmological constant might prevent such a violation.     
\end{abstract}

\section{Entropy bounds}
The entropy associated to a black hole is given by the formula
\begin{equation}
S_{BH}=\frac{{\cal A}}{4G}\,,\label{bh}
\end{equation}
 where ${\cal A}$ is the horizon area \cite{formula1}. 
 The statistical origin of $S_{BH}$ is still not completely clear. However, its status as thermodynamical entropy is strongly supported by classical and semiclassical  calculations in black hole backgrounds \cite{wald}. These give support to the generalized second law of thermodynamics, which is the ordinary second law but where the black holes are considered as objects having the entropy (\ref{bh}) \cite{bekenstein1}. 

As a result, the association of area and entropy permeates to different situations including flat space ones. Consider for example the following thought experiment \cite{susskind}. Take a nearly flat region of space-time and imagine a spherical shell of matter collapsing over it in such a way to produce a black hole of area ${\cal A}$. Thus, if the generalized second law is right (and if the black hole really forms \cite{wald,forms}) the original entropy present in the spherical region $V$ of boundary area ${\cal A}_V={\cal A}$ satisfies the spherical bound \cite{bousso} 
 \begin{equation}
 S(V)\le\frac{{\cal A}_V}{4G}\,.\label{bo}
\end{equation}
This may indicate that the dimension of the Hilbert space of physical states which can be localized in $V$  is finite and given by $e^{\frac{{\cal A}_V}{4G}}$ \cite{thooft}. In that case any entropy $S(W)$ for a region $W\subseteq V$ would also be bounded by ${\cal A}_V/(4G)$. This is not simply a consequence of (\ref{bo}) since in general the entropy is not monotonously increasing with size. This stronger form of the spherical bound can be  called the holographic bound.

Different entropy bounds suggested by black hole thermodynamics have been proposed, and some covariant forms have been developed which are also applicable to regimes of strong gravity.
In particular, the Bousso entropy bound has (\ref{bo}) as a special case \cite{bousso1}. It involves the so called light-sheets. Given a two dimensional spatial surface $\omega$, a light-sheet $H_\omega$ for it is a null hypersurface orthogonal to $\omega$ which has everywhere non-positive expansion\footnote{The non-positive expansion condition is not completely mysterious from the microscopic point of view, since it arises naturally  from the compatibility between the covariance and simple model of a cutoff for the number of degrees of freedom  \cite{hc1}. }. The bound reads
\begin{equation}
S(H_\omega)\le \frac{{\cal A}_\omega}{4G}\,.\label{bousso} 
\end{equation}
Similarly, if a light-sheet $H_{\omega_1}^{\omega_2}$ starts at a spatial two dimensional surface $\omega_1$ and ends at a spatial surface of smaller area $\omega_2$, the FMW bound reads \cite{fmw}
\begin{equation}
S(H_{\omega_1}^{\omega_2})\le \frac{{\cal A}_{\omega_1}-{\cal A}_{\omega_2}}{4G}\,.\label{mfw}
\end{equation} 
This is stronger than the Bousso bound and implies both, the generalized second law and the Bekenstein bound \cite{bekenstein}. 
 
The bounds (\ref{bousso}) and (\ref{mfw}) hold in classical regimes, with the assumption that the entropy is extensive and can be computed out of some sort of entropy density \cite{fmw,otro}. It is also required that the entropy density is bounded above by an expression involving the energy density and the Planck mass. In this case the reason behind the validity of the bounds turns out to be simply that the entropy requires energy, which in turn curves the space in such a way that (\ref{bousso}) and (\ref{mfw}) hold (for example by ending a light sheet). Thus, classical gravity is what keeps the bounds safe according to these examples in the classical regime \cite{bousso}.
    
\section{Entanglement entropy}
Here we are interested in testing the entropy bounds  in empty flat space. Classically the entropy bounds are safe in this case, since no entropy is present. Beyond the classical regime however, the vacuum fluctuations start to play a very relevant role. 

Indeed, the quantum state $\rho_V$ relevant to the algebra of operators acting on a local region $V$ follows from the trace of  the vacuum state over the exterior region Hilbert space ${\cal H}_{-V}$,
\begin{equation}
\rho_V=\textrm{tr}_{{\cal H}_{-V}}\left| 0\right> \left< 0 \right|\,.
\end{equation}
The corresponding entropy (entanglement entropy) is given by
\begin{equation}
S(V)=-\textrm{tr} \rho_V \log(\rho_V)\,,
\end{equation}
and is divergent \cite{bombelli}.

The divergent terms are proportional to quantities which are local in the boundary of $V$. That is, in four dimensions we have \cite{ch1}
 \begin{equation}
 S(V)=g_2[\partial V] \,\epsilon^{-2} + g_1[\partial V]\,\epsilon^{-1} + g_0[\partial V]\,\log (\epsilon)+ S_0(V)\,,   \label{div}
 \end{equation}
 where $S_0(V)$ is a finite part, $\epsilon$ is a short distance cutoff, and the $g_i$ are local and extensive  functions on the boundary $\partial V$, which are homogeneous of degree $i$. The leading divergent term coefficient  $g_2[\partial V]$ is proportional to the square of the size of $V$, what is usually referred to as the area law for the entanglement entropy \cite{srednicki}. However, $g_2$ and $g_1$ depend on the regularization procedure and $g_2$ is not proportional to the area if this later is not rotational invariant. These two terms are not physical within QFT since they are not related to continuum quantities. In a series of recent papers, this type of expansion was also found and exploited in the context of the AdS-CFT duality \cite{ryu}.
      
The particular form of the divergent terms in (\ref{div}) is due to both, the local nature of the ultraviolet divergences, and that the boundary is shared between $V$ and $-V$ which have the same entropy (for any pure global state)
\begin{equation}
S(V)=S(-V)\,.
\end{equation}
On a technical level it is also the consequence of the fact that the entanglement entropy is the variation of the free energy under small conical singularities located on the boundary of $V$ \cite{tech,fursaev}. 
Physically, the divergences have their origin in the same problem which obstructed the realization of a relativistic quantum mechanics of a single particle. The localization of the particle in a size smaller than its mass gives place to the creation of particle-antiparticle pairs, what in turn requires the simultaneous description of any number of particles. The localization in $V$ also produces pairs of particles, one member going inside $V$ and the other outside it, leading to an infinite amount of entanglement. 
 This divergent entropy must be present in the localized entropy for any state, and also in curved space. 
 
 In contrast, the finite value for the entropy of relativistic gases refers to the infinite volume limit. In that case one usually considers the theory at a finite temperature to be defined in a box with some specific boundary conditions,  avoiding the entanglement with the exterior region. Any boundary condition for the finite volume theory, or a cutoff for the localized state in the unbounded theory, give the same result for the entropy per unit volume in the infinite volume limit, since the differences appear as area growing terms.   
However, the ambiguities are unavoidable when we need a prescription for calculating the entropy in a finite region, such as a black hole, the region inside a cosmological horizon, or the flat space region $V$ in the bound (\ref{bo}).

In fact, the same divergences show up in the QFT calculation of the black hole entropy, where the horizon replaces the artificial division of inside and outside given by the boundary of $V$ \cite{bombelli}. According to a widespread view, the finite result (\ref{bh}) would be the manifestation of the quantum nature of space-time. Quantum gravity would cutoff the ordinary QFT calculation below a certain size of order of the Planck length, where the description of space-time in terms of a classical manifold breaks down. In this scenario it is expected that a fundamental cutoff regulates also the flat space entropy \cite{suss}.

\section{Mutual information} 
In absence of a specific prescription for the physical cutoff on the entanglement entropy we have to deal with the QFT results. 
 In analogy with the calculation of the Casimir energy, the worst effects of the vacuum fluctuations may be eliminated, but not at a zero cost. The subtraction of divergences must be done carefully and the entanglement entropy of the vacuum cannot be simply put to zero. This is because there are finite quantities which are universal (independent of regularization prescription) and which can be derived from it. One such quantity is the mutual information \cite{ch2,ch3,ch4} 
\begin{equation}
I(V,W)=S(V)+S(W)-S(V \cup W) \label{mutual}
\end{equation}
between two non intersecting sets $V$ and $W$.  Note that all divergent terms get subtracted in $I(V,W)$\footnote{The entanglement entropy for gauge fields and non-minimally coupled scalars contain a negative contribution in the form of contact terms which could violate the positivity of the entropy \cite{gauge,demers}. However, these are harmless to the physical quantity $I(V,W)$, since the contact terms also get subtracted away.}. The mutual information is used in statistics as a measure of the information shared by two systems. In the present context it can be thought as a correlator between non local objects, which may be defined for any QFT irrespective of the field content. 
In special cases one can compute it explicitly.  For large separating distances between $V$ and $W$ it behaves as expected, is exponentially decreasing for massive theories and decreasing as a power of the distance for massless ones. 
We remark that $I(V,W)$ is a well defined quantity at the axiomatic level in QFT (using an alternative expression rather than (\ref{mutual})), with the only requirement that there must be a non zero separation distance between $V$ and $W$ \cite{ch2,araki}.  

In fact, imagine that $W$ gets larger and larger covering $-V$, and the boundary of $W$ approaches the one of $V$ from outside, encircling it. In the limit when $V$ and $W$ cover the hole space we would have heuristically $S(V)\sim S(-W)=S(W)$, $S(V\cup W)\sim 0$ and 
\begin{equation}
I(V,W)\sim 2 S(V)\,.\label{fa}
\end{equation}
 Thus, the mutual information diverges as the boundaries of the sets get closer to each other, somehow reproducing the divergences which were present in $S(V)$, but now in a universal way. In this sense $I(V,W)$ is a kind of "point splitting" regularization of $S(V)$. 

The form of the leading term in the mutual information for two sets $V$ and $W$ approaching to each other in a conformal field theory can be inferred from dimensional analysis. If two parallel faces of area ${\cal A}$, one from $V$ and one from $W$, are at small distance $d$, we have in the conformal case
\begin{equation}
I(V,W)\simeq \kappa\, \frac{{\cal A}}{d^{n-2}}\,,\label{tuy}
\end{equation}
where $n$ is the space-time dimension \cite{ch4}. We have computed explicitly the dimensionless constant $\kappa$
for free fields 
\begin{equation}
\kappa =  \frac{\textrm{vol}(S^{n-3})}{(n-2)(2\pi)^{n-2}}\int_0^\infty dt \,t^{n-3}\, (n_B c_B(t)+n_F c_F(t)) \,,\label{ttt}
\end{equation}
where $n_B$ and $n_F$ are the multiplicity of the bosonic and fermionic degree of freedom, $\textrm{vol}(S^{n-3})$ is the volume of a $(n-3)$ dimensional unit sphere, and $c_B(t)$ and $c_F(t)$ are the two dimensional entropic functions corresponding to boson and fermion fields \cite{ch3,ch4}. These are defined by the formula $c(mL)=L\,dS(L)/dL$, where $S(L)$ is the entanglement entropy corresponding to an interval of length $L$ in $1+1$ dimensions \cite{ch2}. The subindex $B$ corresponds to a two dimensional real scalar field and $F$ to a Majorana fermion one, and $m$ is the field mass. We have in four dimensions the approximate result
\begin{equation}
\kappa= 0.0055 \, n_B+0.0053\, n_F\,.
\end{equation}

The mutual information $I(V,W)$ can be thought as  the localized form of entropy which replaces $S(V)$ in the QFT \cite{futuro}. It is also a positive quantity, and in addition, it has a very nice property which does not hold for the entropy: it is monotonically increasing with the size of the sets \cite{stat},
\begin{equation}
 I(V,W)\le I(V,U), \hspace{2.5cm} W\subseteq U \,.\label{ine1}
\end{equation} 
This means that the mutual information is smoothly varying with the set, and that the sharpness in the definition of  the set boundaries is not very relevant to it. 

Using the inequality (\ref{ine1}) and the fact that it is always possible to consider the density matrix $\rho_{V\cup W}$ as a partial trace of a pure state on a bigger space, we can show   
\begin{equation}
I(V,W)\le 2 \min(S(V),S(W))\,.\label{fofo}
\end{equation}  
These mathematical properties of  $I(V,W)$ are reasonable, and make justice to the name of mutual information.

\section{The FMW bound in Minkowski space}
The formula (\ref{fofo}) is the starting point of our application of the mutual information to the entropy bounds. The mutual information $I(V,W)$ is calculable in QFT and in principle it might help to state and test the entropy bounds beyond the classical regime. In first place we note that the combination of (\ref{fofo}) with the actual results in QFT (see for example formula (\ref{tuy})) show that the entanglement entropy in flat space,  regularized by a cutoff at high energy, cannot be renormalized to zero. In general we expect it to be very large, if actually finite (see also the discussion at the end of \cite{tech}).  This is because we can use the QFT safely at least up to separating distances which are large with respect to the Planck length (we elaborate more on this delicate point below). This big amount of entropy may pose difficulties to the entropy bounds, and at the same time make them more constraining and interesting.

In particular, consider the case of the FMW bound in Minkowski space. It gives zero entropy crossing for pieces of the null hyperplanes since the bounding areas $\omega_1$ and $\omega_2$ are equal in this case. 
 Imagine then a polyhedron $P$ included in a spatial hyperplane in flat space.  It has a Cauchy surface formed by several null surfaces $H_i$ included in null hyperplanes. Each $H_i$ corresponds to one of the polyhedron faces, to which it is orthogonal. The total entropy has to be less than the sum of the entropies of the individual $H_i$ since the entropy is subadditive. Thus we have according to the FMW bound
\begin{equation}
S(P)\le \sum_i S(H_i)=0\,. 
\end{equation}     
 However, $P$ has a non zero (and in general very large) entropy due to the vacuum fluctuations. 
 
We could try to reinterpret the FMW bound in such a way that this counterexample would not work. A possibility is to ask eq. (\ref{mfw}) to hold for the difference in the entropies crossing $H_{\omega_1}^{\omega_2}$ for a given global state and the vacuum state. This subtraction resembles the calculations done in the refs. \cite{works} in relation to the Bekenstein bound (which is related to the FMW bound). The problem is then to formulate this new version in a fully relativistic manner.

\section{The holographic bound}
Consider now the holographic bound for a sphere $V$. It reads in terms of the mutual information
\begin{equation}
I(U,W)\le \frac{{\cal A}_V}{2G}\,\label{bonu}
\end{equation}
for any $U\subseteq V$ and any $W$ nonintersecting with $U$.
The case of a set $W$ with boundary approaching the one of $V$ shows that this bound is always violated naively in QFT, since when the separating distance goes to zero $I(V,W)$ tends to infinity. However, to violate the bound in this way we need to use Planck scale distances between $V$ and $W$. The space-time structure is uncertain at this scale, and the QFT assumption presumably breaks down.

\begin{figure}[t]
\centering
\leavevmode
\epsfysize=6cm
\epsfbox{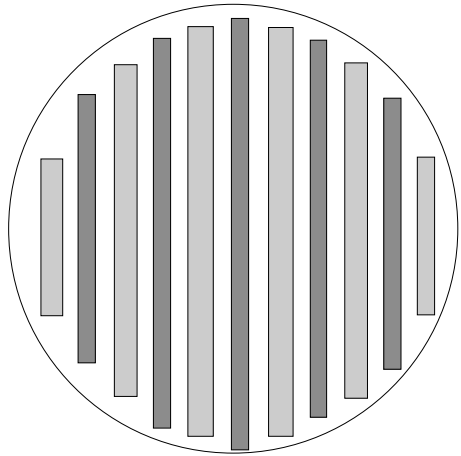}
\caption{The sets $U$ (light shaded) and $W$ (dark shaded) are nonintersecting and have   large bounding area inside the sphere $V$.}
\end{figure}

Then, consider instead the following construction. Let $U$ be a set included in $V$ with  large bounding area ${\cal A}_U\gg {\cal A}_V$. Let also $W$ be another set with boundary approximately parallel to the one of $U$, that is, located at a distance from it which is kept around a fixed value $d$ (i.e. the multi-layered sets in figure 1). The area ${\cal A}_U$ is then bounded as
\begin{equation}
{\cal A}_U\lesssim {\cal A}_V \frac{R_V}{d}\,,
\end{equation}
where $R_V$ is the sphere radius. 
According to eq. (\ref{tuy}) the mutual information between $U$ and $W$ can be very large,
\begin{equation}
I(U,W)\sim \frac{{\cal A}_U}{d^2}\lesssim \left(\frac{R_V}{d}\right)^3 \,.\label{volume}
\end{equation}
If one chooses
\begin{equation}
R_V\gtrsim M_{Pl}^2\, d^3\,,\label{holds}
\end{equation}
with $M_{Pl}$ the Planck mass, the mutual information $I(U,W)$ 
exceeds the holographic bound ${\cal A}_V/(2G)$, even if the separating distance $d$ is large in Planck units. 

The equation (\ref{volume}) gives a cutoff independent (and infinite volume limit independent) meaning to the statement that the number of degrees of freedom in a QFT grows as the volume. Thus, the "area law" for the entanglement entropy in QFT does not refer to the number of internal degrees of freedom. It rather means simply that the number of degrees of freedom which are connected with the exterior grows as the area.   

On the other hand, once gravity is turned on one encounters objections to the application of the QFT results even in this case involving distances which are large in Planck units. The field modes through which the information is shared between $U$ and $W$ are of typical wavelength $d$, and there are ${\cal A}_U/d^2$ of these modes. 
If (\ref{holds}) holds, the typical energy involved in the vacuum fluctuations is then of order 
\begin{equation}
E\sim \frac{{\cal A}_U}{d^3}\gtrsim \frac{{\cal A}_V}{2Gd}\sim \frac{{R_V}}{d}\, (M_{Pl}^2 R_V)\,,
\end{equation}
which is greater than the mass of a black hole of radius $R_V$.
Then, in principle it is possible that quantum gravity reduces the mutual information in this example through (virtual) black holes \cite{thooft}.
 Naively one could have expected otherwise, that the gravitons add to the mutual information rather than diminish it\footnote{However, the amount of information the gravitons can carry is unclear even at the classical level  \cite{f}}.
 
Note that in any case, the physical "ultraviolet" cutoff must act paradoxically in a non local way in order to prevent the violation of the holographic bound. A covariant local cutoff to the entropy (such as a Pauli-Villars regularization \cite{demers}) is not enough. 

\section{The Bousso bound and conformal theories}
On the other hand, the spherical bound and its generalization, the Bousso bound, refer to the entropy, not to the number of degree of freedom, and are not in conflict with the construction above.  

The question is then if it is possible to violate this bound using (\ref{fofo}) in a situation where the sets involved are at a distance to each other which is large in Planck units. This is not possible for sets lying in the same spatial hyperplane (for a reasonably small number of fields), but, strangely enough, can be done by locating  them in space-time. 

\begin{figure}[t]
\centering
\leavevmode
\epsfysize=6cm
\epsfbox{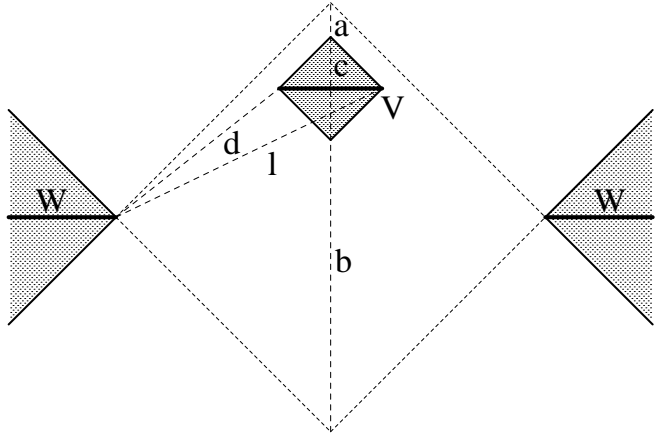}
\caption{Time corresponds to the vertical axes and the horizontal axes is one of the spatial coordinates. The full plot is spherically symmetric. The thick horizontal lines represent the spatial sets $V$ and $W$. The shaded regions bounded by solid lines are their domain of dependence (causal shadow). The diameter of $V$ is $c$ while the interior diameter of $W$ is $a+b+c$. The set $W$ extends to spatial infinity.}
\end{figure}

In order to  have large mutual information at long separation distances we consider a conformal field theory.  Take $V$ to be a sphere. Since the mutual information is increasing under inclusion, we chose the set $W$ to be as large as possible being at a given distance from $V$, as shown in the figure (2). The conformal invariance of the vacuum state implies that the mutual information for this geometry is a function of the cross ratio 
\begin{equation}
I(V,W)= f(\eta)\,, \hspace{1.7cm}\eta=\frac{a\, b}{(a+c)\,(b+c)}\,, \label{eta}
\end{equation}
with $0\le \eta\le 1$. 
Then, in order to compute $I(V,W)$, we can put $V$ and $W$ in the same spatial plane, and if $\eta$ is small we have $a=b\ll c$. In this case the thin annulus between $V$ and $W$ tends to a long strip, and we use equation (\ref{tuy}) to show that   
\begin{equation}
f(\eta)\sim \frac{\pi\,\kappa}{\eta}\,, \hspace{1.5cm} \eta \ll 1\,.
\end{equation}

Now consider the case $a\ll c\ll b$. We have in this case  $\eta\simeq a/c\ll1$ and 
\begin{equation}
I(V,W)= \pi\kappa\ \frac{c}{a}=\kappa \frac{\sqrt{{\cal A}_V {\cal A}_W}}{d^2}\,, \label{esi}
\end{equation}
where $d=\sqrt{a\,b}$ is the spatial distance between the set edges. 

Thus, given $V$ we can keep $d$ fixed to any given distance and make the area of $W$ grow indefinitely
in order to surpass the bound for $V$. To do so according to (\ref{bo}) and (\ref{fofo}) we need 
\begin{equation}
\frac{M_{Pl}^2}{2\kappa} \, d^2\, R_V \lesssim R_W\,, \label{tara}
\end{equation}
where $R_V=c/2$ and $R_W\sim b/2$ are the radius of the spheres corresponding to the boundaries of $V$ and $W$. 
This inequality can be satisfied for $R_W$ large enough, without requiring $R_V$ or $d$ to be short\footnote{The same type of gathering of entanglement in a small set occurs in black holes \cite{futuro}. The mutual information can be used as a tool to determine the localization of the entanglement with the Hawking radiation inside the black hole, and it seems this later is correlated mostly with the neighborhood of the initial point from which the horizon emanates.}.

However, as in the case of the holographic bound previously discussed, this violation of the Bousso bound is not free from its own internal difficulties. 
For example, there are sub-Planckian scales in this construction, such as $a$. This is what invalidates the arguments which lead to (\ref{esi}) in the case we model the theory in a square lattice of Planck length spacing. In this case the existence of a bound on the entropy (different from (\ref{bo})) for a finite volume set is evident. However, the physical cutoff should respect long distance covariance, and we could have used for $\eta$ the equivalent expression  $\eta=d^2/l^2$ (instead of (\ref{eta})), which does not involve any small distances. 

Also, the field modes which interchange information between the sets are of large   angular momentum $L\sim \sqrt{R_V\, R_W}/d$. This can be seen by transforming back to the situation where the sets are in the same spatial hyperplane, where these modes have typical wavelength corresponding to the separation distance between $V$ and $W$. This angular momentum gives one mode per solid angle $L^{-2}\sim d^2/(R_V\, R_W)\lesssim (A_V M_{Pl}^2)^{-1}$, what  represents more than one mode per Planck area on the surface of $V$.
However, a direct angular discreteness is not compatible with large scale rotational invariance. This problem points out again at least an unusual way the physical cutoff is required to act to save the bound.   

Maybe there is a hint in eq. (\ref{tara}) of a different way gravity can save the bound here. In fact, we do not have access to arbitrarily large flat space-times if the cosmological constant is present, what can indicate that we should take $R_W\lesssim R_\Lambda=\sqrt{3/\Lambda}$ in order to apply the arguments above.  Then, equation (\ref{tara}) resembles the one proposed by Dirac in 1937 as an intriguing relation between the Planck scale, the cosmological scales and some scale of particle physics \cite{weinberg}. 
Indeed, at low energies there is only the massless electromagnetic field (we do not consider the graviton). However, due to the  coupling with the electron, its self-interaction is important at energies higher than the electron mass. 
Thus, the distances $d$ and $R_V$ cannot be smaller than a typical size given by electrodynamics where the interactions between fotons break conformal invariance. Let us take heuristically $d^2\,R_V\sim (\alpha/m_e )^3$. The number of powers of $\alpha$ can be different and we ignore the optimal numerical prefactor, which is not necessarily near $1$. This gives an approximate bound 
\begin{equation}
R_\Lambda\lesssim \frac{\alpha^3}{2\kappa}\frac{M_{Pl}^2}{(m_e )^3} =4. \,10^{27}\, \textrm{meter} \,,\label{dirac}
\end{equation}
which is surprisingly of the correct order of magnitude of the cosmological observations \cite{observations},
\begin{equation}
R_\Lambda\sim 1.5 \,10^{26} \,\textrm{meter}\,.
\end{equation}
A massless neutrino can surpass the bound, since it has much smaller interactions. This idea needs to be further explored using QED in a de Sitter background. 

The argument above is connected to a theorem proved in ref. \cite{cas}. There it was shown that if the entanglement entropy is finite and Lorentz invariant it must be exactly proportional to the boundary area. The proof is based on the strong subadditive property of the entropy, and involves crucially short and long distances of arbitrary size in Minkowski space. An entropy proportional to the area means that the mutual information is identically zero, $I(V,W)\equiv 0$. As we have mentioned this simply cannot hold since the mutual information, at least in certain situations,  is a low energy quantity which is non zero in QFT. Of course, QFT resolves this tension just making the entropy divergent. If we insist in keeping the entropy finite and covariant there must be a short distance cutoff invalidating the arguments involving arbitrarily short scales. But it is possible that this is not enough if the theory has a non trivial infrared fixed point. In this case we could also need an infrared scale to keep $I(V,W)$ for, say, human size configuration of sets $V$ and $W$, fixed to a given positive value of order one.

In conclusion we can say that the formulation and testing of the entropy bounds at a semiclassical level require  treating QFT in a bounded region. This means for example imposing a cutoff or an artificial boundary condition. These procedures necessarily give ambiguous values for the entropy because they contain input from external elements. In this sense the mutual information offers a unique way to deal in a cutoff independent way with the localized entropy in QFT. At the same time it maintains certain inequalities with the entropy which allow stating the entropy bounds unambiguously. It clarifies the role of the vacuum fluctuations showing that they lead to violations to the entropy bounds even in empty space. It also shows that if the entropy bounds hold in nature some essentially non local effects must be produced by the physical cutoff, even in very simple and specific flat empty space situations. These could be simpler scenarios that black holes in order to contrast ideas.

\section{Acknowledgments}
I thank Marina Huerta for discussions and for long standing collaboration, on the results of which this paper is based.

\end{document}